\documentclass[twocolumn,showpacs,prl]{revtex4}

\usepackage{mathrsfs}
\usepackage{amsmath,amssymb}
\usepackage{graphicx}
\usepackage{verbatim}

\draft
\begin{document}

\title{Andreev Bound states in One-Dimensional Topological Superconductor}

\author{Xiong-Jun Liu}

\affiliation{Joint Quantum Institute and Condensed Matter Theory Center, Department of Physics,
University of Maryland, College Park, Maryland 20742, USA}

\begin{abstract}
We study the charge character of the Andreev bound states (ABSs) in one-dimensional topological superconductors with spatial inversion symmetry (SIS) breaking. Despite the absence of the SIS, we show a hidden symmetry for the Bogoliubov de Gennes equations around Fermi points in addition to the particle-hole symmetry. This hidden symmetry protects that the charge of the ABSs is solely dependent on the corresponding Fermi velocities. On the other hand, if the SIS is present, the ABSs are charge neutral, similar to Majorana fermions. We demonstrate that the charge of the ABSs can be experimentally measured in the tunneling transport spectroscopy from the resonant differential tunneling conductance.
\end{abstract}
\date{\today}
\maketitle

\textit{Introduction -} In superconductors (SCs), Bogoliubov quasiparticle (BQ) is a coherent superposition of the electron and hole; hence the charge carried by a BQ is not conserved \cite{Schrieffer}. The effective charge of the BQ can be identified as the expectation value of the charge operator. Based on this definition, the charge of the quasiparticle excitation above SC gap is generically a continuous function of the energy and SC order parameter, and therefore it is expected to be sensitive to perturbations \cite{Schrieffer}.

A peculiar subgap quasiparticle state in topological SCs is Majorana fermion (MF), which is charge neutral and equal to its own antiparicle. Driven by the exotic properties that MFs obey non-Abelian statistics and have potential applications in fault tolerant topological quantum computation \cite{Ivanov,Sankar,Nayak}, the search for MFs in realistic condensed matter systems has become an exciting pursuit. Candidates shown to exhibit topological superconductivity and host MFs include two dimensional (2D) $p+ip$ \cite{Read,Lee,Luke,Kwon,Tewari,Sato} and 1D $p_x$-wave SCs \cite{Sengupta,Kitaev1}, topological insulator/SC heterostructures \cite{Fu1,Nilsson,Fu2,Linder}, and spin-orbit (SO) coupled semiconductor/SC heterostructures \cite{Sau,Alicea,Roman0,Roman,Oreg,Potter,Alicea1}. MFs are self-Hermitian and therefore charge neutral, which in SC is guaranteed by the particle-hole symmetry \cite{Read}. Since the self-Hermitian property exists only for zero modes, a natural question arises: what is the charge of a generic (non-Majorana) Andreev bound state (ABS)? Furthermore, in the tunneling transport spectroscopy it is predicted that a single MF can induce the differential tunneling conductance (DTC) peak of height $2e^2/h$ at the resonant Andreev reflection \cite{Law,Flensberg,Wimmer}, which is a consequence of the  self-Hermitian and charge neutrality of MFs. Is such peak value of DTC unique to the MFs? These fundamental questions motivate us to look more closely at the ABSs in SCs.

In this Letter we focus on the ABSs at Josephson junction of 1D topological SCs. We show a hidden symmetry for the Bogoliubov-De Gennes (BdG) equations around Fermi energy, by which we predict that the charge of ABSs solely depends on the Fermi velocities, regardless of other details. We also demonstrate that the charge can be measured by DTC at the resonant Andreev reflection.

{\textit{Generic theory -}} We start from the generic mean-field Hamiltonian for 1D spinless p-wave SC \cite{Read,Kitaev1}:
$H=\sum_{k}({\cal E}_{k}-\mu)c^\dag_{k}c_{k}+\int dx\Delta(x)c(x)c(x+a)+h.c.$, where $\Delta$ is the induced SC order parameter, $\mu$ is the chemical potential, $a$ is the lattice constant, $c$ and $c^\dag$ are annihilation and creation operators of electrons, respectively. A broken spatial inversion symmetry (SIS) is generically allowed for the present study and is characterized by the single electron dispersion relation ${\cal E}_{k}\neq{\cal E}_{-k}$, which leads to a discrepancy between Fermi velocities $v_{fL}$ and $v_{fR}$ for the left and right movers. Different mechanisms for SIS breaking will be studied later in lattice model and in semiconductor nanowire systems.

In the parameter regime that $|\Delta|\ll E_F$ with $E_F$ the Fermi energy measured from the bottom of the band, we can expand the Hamiltonian around the Fermi points by the transformation $c(x)=c_{R}(x)e^{ik_{R}x}+c_{L}(x)e^{-ik_{L}x}$, and keep the terms up to the first order of $1/(k_{R,L}\xi)$ with $\xi$ the coherence length of the SC. Here $k_{L}$ and $k_R$ are respectively Fermi momenta at left and right Fermi points. With the help of the Nambu bases ${\cal \psi}_+=[c_{R}(x),c^\dag_{L}(x)]^T$ and ${\cal \psi}_-=[c_{L}(x),c^\dag_{R}(x)]^T$, we rewrite the Hamiltonian in explicit particle-hole symmetric form $H=\int dx{\cal \psi}_+^\dag(x){\cal H}_+(x){\cal \psi}_+(x)+\int dx{\cal \psi}_-^\dag(x){\cal H}_-(x){\cal \psi}_-(x)$, with ${\cal H}_-(x)$ related to ${\cal H}_+(x)$ by particle-hole transformation ${\cal H}_-(x)=-\tau_y{\cal H}^*_+(x)\tau_y$, and
\begin{eqnarray}\label{eqn:H2}
{\cal H}_+(x)={\left[
\begin{matrix}
-i v_{fR}\partial_x & i\tilde{\Delta}(x)\\
-i\tilde{\Delta}^*(x) & i v_{fL}\partial_x\end{matrix} \right]}.
\end{eqnarray}
The explicit form of $\tilde{\Delta}(x)$ depends on the electron dispersion relation ${\cal E}_{k}$ and Fermi momenta, but as shown below, it does not affect the charge of the ABSs.

The ABSs can be studied by considering a Josephson junction formed around $x=0$. We denote by the ABS wave
functions $\Phi_{\nu\pm}(x)=[u_{\nu\pm}(x), v_{\nu\pm}(x)]^T$ for the BdG Hamiltonians ${\cal H}_\pm$ with eigenvalues ${\cal E}_{\nu\pm}$ $(\nu=...-1,0,1...)$. The Bogoliubov
operators of the ABSs are then defined as $b_{\nu\pm}=\int dx[u_{\nu\pm}(x)c_{R/L}(x)+v_{\nu\pm}(x)c_{L/R}^\dag(x)]$ 
and the electric charges are calculated by $e^*_{\nu\pm}=e\int dx[|u_{\nu\pm}(x)|^2-|v_{\nu\pm}(x)|^2]$. However, the wave functions of ABSs are not analytically solvable in the generic situation. Fortunately, without solving the wave functions, we shall show that $u_{\nu\pm}(x)$ and $v_{\nu\pm}(x)$ satisfy a universal relation for all ABSs: $|u_{\nu+}(x)/v_{\nu+}(x)|=|u_{\nu-}(x)/v_{\nu-}(x)|^{-1}=(v_{fL}/v_{fR})^{1/2}$. We then have
\begin{eqnarray}\label{eqn:charge2}
e^*_{\nu\pm}\equiv e^*_\pm=\pm e\bigr(\frac{v_{fL}-v_{fR}}{v_{fR}+v_{fL}}\bigr).
\end{eqnarray}

We prove the above result with the BdG Hamiltonian ${\cal H}_{+}$. The SC order parameter is written as $\tilde{\Delta}(x)=\Delta_1(x)+i\Delta_2(x)$. Our main goal is to show that the relation $v_{fR}|u_{\nu+}(x)|^2=v_{fL}|v_{\nu+}(x)|^2$
is valid for all ABSs in generic case. For this purpose we define $G_{\nu+}(x)=|u_{\nu+}(x)/v_{\nu+}(x)|^2$, and derive from BdG equations that
\begin{eqnarray}\label{eqn:general6}
\frac{dG_{\nu+}(x)}{dx}=2\bigr(\frac{|v_{\nu+}(x)|^2}{v_{fR}}-\frac{|u_{\nu+}(x)|^2}{v_{fL}}\bigr)\frac{\mbox{Im}(\tilde{\Delta} u^*_{\nu+} v_{\nu+})}{|v|^2}.
\end{eqnarray}
Note that $\mbox{Im}(2\tilde{\Delta} u^*_{\nu+} v_{\nu+})=v_{fR} d|u_{\nu+}|^2/dx\neq0$ for bound states. Then if one has $\frac{dG_{\nu+}(x)}{dx}=0$, the only solution is $|u_{\nu+}(x)/v_{\nu+}(x)|^2=v_{fL}/v_{fR}$. On the other hand, we notice that $\partial_x f_{\nu+}(\Delta_1,\Delta_2)=0$, where $f_{\nu+}(\Delta_1,\Delta_2)=v_{fR}|u_{\nu+}|^2-v_{fL}|v_{\nu+}|^2$. Therefore the function $f_{\nu+}(\Delta_1,\Delta_2)$ is independent of position.
To determine the value of $f_{\nu+}(\Delta_1,\Delta_2)$ we examine the asymptotic behavior of the ABS wave function. We take the boundary condition that the order parameter $\tilde{\Delta}(x\rightarrow\pm\infty)$ is finite. The asymptotic behavior of $u_{\nu+}(x)$ and $v_{\nu+}(x)$ can be generically described by $u_{\nu+}(x\rightarrow+\infty)\sim\frac{A_1}{x^{\delta_1}}e^{-\gamma_1x+i\tilde{\gamma}_1x}$ and $v_{\nu+}(x\rightarrow+\infty)\sim\frac{A_2}{x^{\delta_2}}e^{-\gamma_2x+i\tilde{\gamma}_2x}$, respectively. By examining the BdG equations of $u_{\nu+}(x)$ and $v_{\nu+}(x)$ at $x\rightarrow+\infty$ one can verify that the exponents must satisfy the relations $\delta_1=\delta_2$ and $\gamma_1=\gamma_2$.
This confirms that $G_{\nu+}(x\rightarrow+\infty)$ is a constant; hence $\frac{dG_{\nu+}(x\rightarrow+\infty)}{dx}\equiv0$. According to Eq.~\eqref{eqn:general6} it follows then $G_{\nu+}(x\rightarrow\infty)=v_{fL}/v_{fR}$. Thus we get $f_{\nu+}(\Delta_1,\Delta_2)=0$, which is valid for the entire position space. Finally we reach the result $v_{fR}|u_{\nu+}(x)|^2=v_{fL}|v_{\nu+}(x)|^2$, completing the proof. Eq.~\eqref{eqn:charge2} gives the charge carried by any ABS in the generic case. We emphasize that this proof is valid for any bound state in the bulk region (with wave function exponentially vanishing for $x\rightarrow\pm\infty$).

The results of the charges $e_\pm^*$ indicate a hidden symmetry for the BdG Hamiltonians. Let $\Phi_{\nu\pm}=[u_{\nu\pm}(x), v_{\nu\pm}(x)]^T$ satisfy ${\cal H}_\pm\Phi_{\nu\pm}={\cal E}_{\nu\pm}\Phi_{\nu\pm}$. We denote by $\mathcal{K}$ and $\mathcal{U}_{{\cal E}_{\nu\pm}}=e^{i{\cal E}_{\nu\pm}(\frac{1}{v_{fR}}-\frac{1}{v_{fL}})x}$ the complex conjugate and the U(1) unitary transformations, respectively. Furthermore, we define the Lorentz boosts $\mathcal{L}_\pm=e^{\lambda_\pm\tau_z}$ to rescale $u_{\nu\pm}(x)$ and $v_{\nu\pm}(x)$, with $\lambda_\pm=\pm\ln(v_L/v_R)^{1/2}$ and $\tau_{x,y,z}$ the Pauli matrices acting on Nambu space. It can be verified that the Hamiltonians are invariant under transformations $\bold\Xi_{\nu\pm}{\cal H}_\pm\bold\Xi^{-1}_{\nu\pm}={\cal H}_\pm$, where $\bold\Xi_{\nu\pm}={U}_{{\cal E}_{\nu\pm}}\mathcal{K}\mathcal\tau_x{\cal L}_\pm$, while the wave functions transform via $\Phi_{\nu\pm}=\bold\Xi_{\nu\pm}\tilde{\Phi}_{\nu\pm}^*$. One can show that the bound states of ${\cal H}_+$ (${\cal H}_-$) are non-degenerate. This leads to $\Phi_{\nu\pm}=e^{i\eta_\pm}\tilde{\Phi}_{\nu\pm}^*$, with $\eta_\pm$ arbitrary constants, which implies that
\begin{eqnarray}\label{eqn:staterelation3}
u_{\nu\pm}(x)=e^{\lambda_\pm} v_{\nu\pm}^{*}(x)e^{i\eta_\pm}e^{i{\cal E}_{\nu\pm}(\frac{1}{v_{fR}}-\frac{1}{v_{fL}})x}.
\end{eqnarray}
This formula recovers the charges $e^*_\pm$ in Eq.~\eqref{eqn:charge2}.
Here we have shown a remarkable result that the electric charge of an ABS is solely determined by the SIS breaking term which is quantified by the discrepancy of the Fermi velocities $v_{f_{L,R}}$, but independent of all other details. This result is protected by the hidden symmetry of the BdG Hamiltonians ${\cal H}_\pm$ under the transformations $\bold\Xi_{\nu\pm}$, and is valid when $1/(k_{L/R}\xi)\ll1$. Particularly, {\em for the inversion symmetric 1D SC  with $v_{fR}=v_{fL}$, the ABSs are charge neutral, regardless of the details of the SC}.

\begin{figure}[b]
\includegraphics[width=1.0\columnwidth]{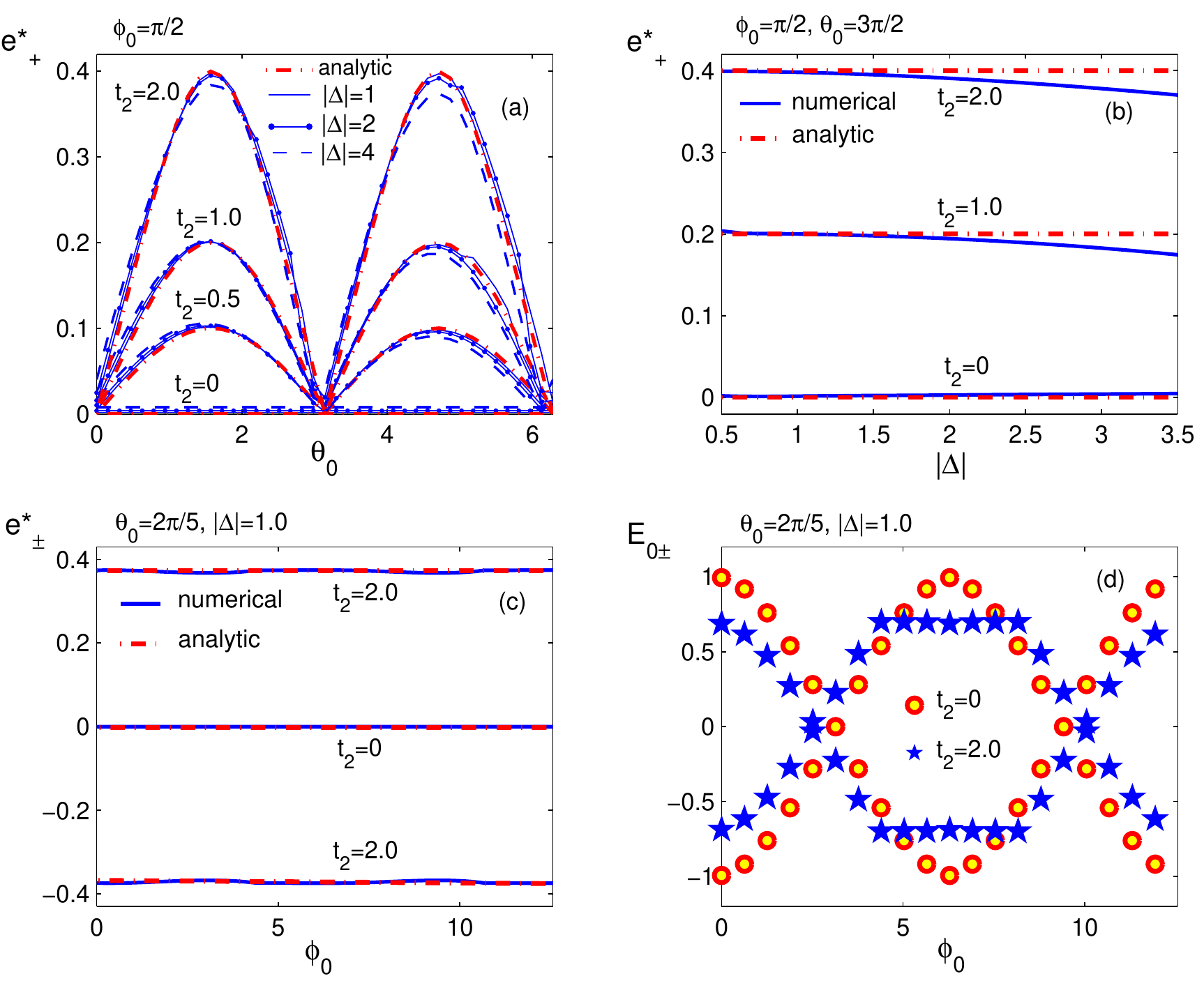}
\caption{(Color online) Numerical (solid lines) and analytic (dash-dotted lines) results for the charge (in unit of $e$) of the ABSs versus $\theta_0$ (a), SC gap $|\Delta|$ (b), and Josephson junction phase difference $\phi_0$ (c), with $t_1$ rescaled to be dimensionless and taken as $t_1=10$. (d) Energy of the ABSs versus $\phi_0$, with $\theta_0=2\pi/5$ and $|\Delta|=1.0$. Other parameters are taken as $\mu\lesssim0$, $t_2=0, 0.5, 1.0, 2.0, \phi_0=\pi/2$ for (a), $t_2=0, 1.0, 2.0, \phi_0=\pi/2,\theta_0=3\pi/2$ for (b), and $t_2=0, 2.0, \theta_0=2\pi/5, |\Delta|=1.0$ for (c-d).} \label{charge}
\end{figure}
{\textit{Lattice Model for spinless p-wave SC -}} Next we study concrete systems with SIS breaking. We consider first the lattice model for 1D topological SC described by
\begin{eqnarray}\label{eqn:H1}
H&=&-\sum_{\langle i,j\rangle}t_1\hat c_i^\dag\hat c_j-\sum_{\langle\langle i,j\rangle\rangle}t_2e^{i\theta_{ij}}\hat c_i^\dag\hat c_j-\mu\sum_i\hat c_i^\dag\hat c_i+\nonumber\\
&&+\sum_{j}(\Delta\hat c_{j}\hat c_{j+1}+h.c.),
\end{eqnarray}
where $t_1$ is the nearest hopping, and the complex next-nearest-neighbor (NNN) hopping $t_2e^{i\theta_{ij}}$ (with $\theta_{i,i\pm2}=\pm\theta_0$) breaks SIS, as in the Haldane model \cite{Haldane} and Kane-Mele model \cite{Kane}. Recent proposals show that the complex NNN hopping with appreciable magnitude may be obtained in the silicene \cite{cliu,silicene}, and realized by doping localized spins in graphene \cite{Hill}. The Hamiltonian Eq.~\eqref{eqn:H1} becomes Kitaev model when $t_2=0$ \cite{Kitaev1}, which can be effectively achieved by e.g., s-wave SC proximity effect in the 1D semiconductor nanowire (NW) with Rashba SO interaction \cite{Roman0,Roman,Oreg,Potter}.
Transforming $H$ into $k$ space yields
$H=\sum_k({\cal E}_k-\mu)c_k^\dag c_k+\sum_k[\Delta\sin(ka)c_{k} c_{-k}+h.c.]$ with ${\cal E}_k=-2t_1\cos(ka)-2t_2\cos(2ka+\theta_0)$.
The electron dispersion relation ${\cal E}_k\neq{\cal E}_{-k}$ (for $k\neq0$) possesses no inversion symmetry when $\theta_0\neq n\pi$ and $|t_2|>0$.

We present in Fig.~\ref{charge} the analytic and numerical results for the ABSs localized on a Josephson junction with phase difference $\phi_0$. Fig.~\ref{charge} (a-b) shows that the numerical solution is well consistent with Eq.~\eqref{eqn:charge2} in the parameter regime that $|\Delta|/E_F\leq1/10$ (equivalent to the condition $1/(k_{L,R}\xi)\leq1/10$). Fig.~\ref{charge} (c-d) confirms that the charge is independent of $\phi_0$ and the energies of ABSs.
It is noteworthy that the crossing point of the ABS spectrum ($\nu=0$) is shifted away from $\phi_0=\pi$ when SIS is broken (blue stars in Fig.~\ref{charge} (d)), while the zero energy crossing itself is topologically stable, at which point two zero MF modes can be introduced \cite{Roman0,Fu2}.

{\textit{Semiconductor Nanowire -}} For the semiconductor NW/SC heterostructure, the SIS is broken when the Zeeman field $(V_x,V_y)=V_0(\cos\theta_0,\sin\theta_0)$ has a tilt angle ($\theta_0$) relative to the NW direction. The Hamiltonian reads
\begin{eqnarray}\label{eqn:nano1}
H&=&\sum_{\sigma,\sigma'=\uparrow,\downarrow}\int dxc_\sigma^\dag(x)\bigr[\frac{\partial_x^2}{2m^*}-\mu+(i\lambda_R\partial_x+V_y)\sigma_y\\
&&+V_x\sigma_x\bigr]_{\sigma\sigma'}c_{\sigma'}(x)+\int dx[\Delta_s(x) c_{\uparrow}(x) c_{\downarrow}(x)+h.c.],\nonumber
\end{eqnarray}
where $m^*,\lambda_R$ and $\Delta_s$ are effective mass of electrons, spin-orbit coupling coefficient, and induced s-wave SC order parameter, respectively. Note that under the spatial inversion transformation $x\rightarrow-x$, one has $(k_x,\sigma_y)\rightarrow(-k_x,-\sigma_y)$, and $\sigma_x\rightarrow\sigma_x$. Therefore the term $V_y\sigma_y$ breaks SIS and results in a discrepancy between Fermi velocities $v_{fL}$ and $v_{fR}$. For the case $\theta_0=0$, the above system is isomorphic to Kitaev model and support Majorana end modes when $V_x^2>|\Delta_s|^2+\mu^2$ \cite{Roman0,Oreg}.

\begin{figure}[tbp]
\includegraphics[width=0.95\columnwidth]{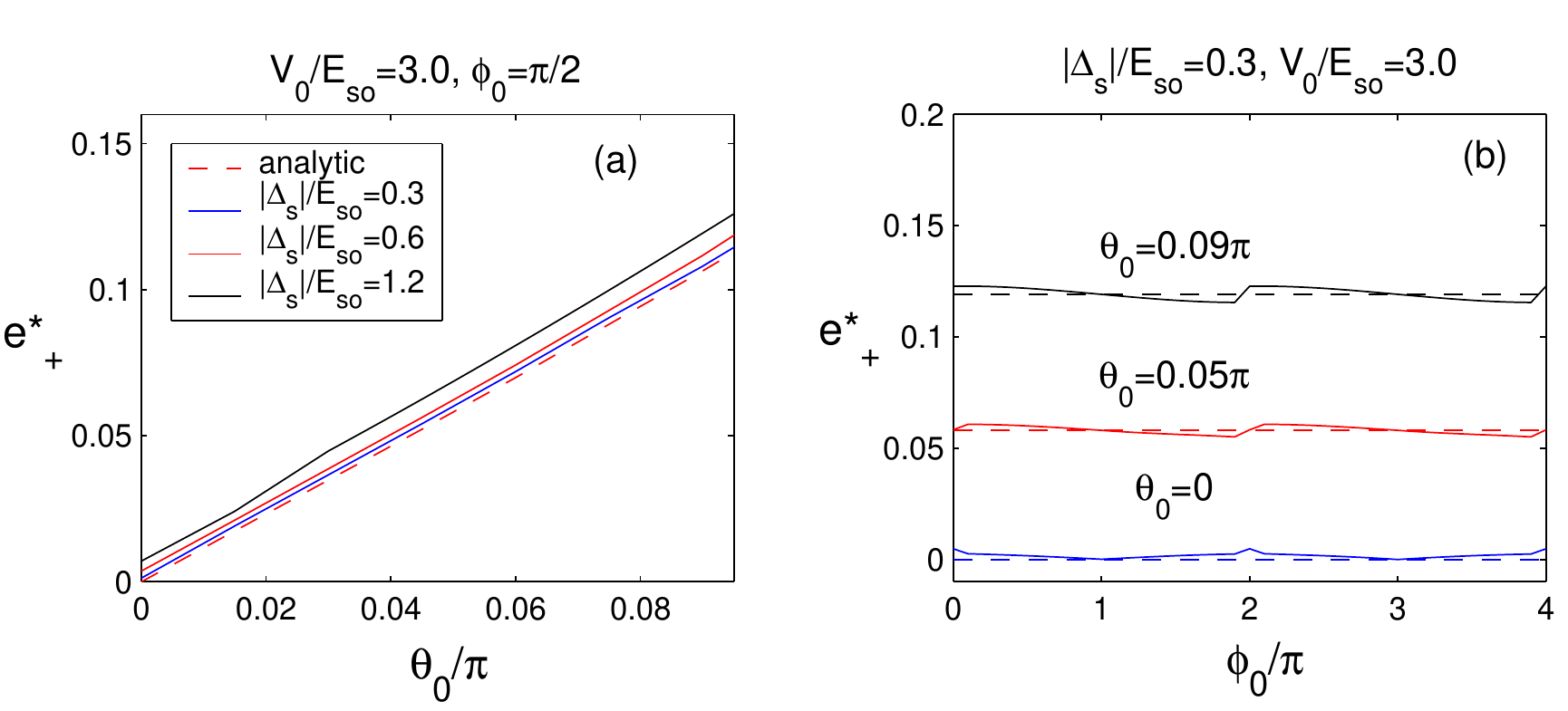}
\caption{(Color online) Numerical (solid curves) and analytic (dashed lines) results for the charge (in unit of $e$) versus tilt angle $\theta_0$ (for (a)) and Josephson junction phase difference $\phi_0$ (for (b)). The spin-orbit coupling energy $E_{so}=m^*\lambda_R^2=1$, and the chemical potential $\mu=0$.} \label{nanowire}
\end{figure}

For a large Zeeman field relative to $|\Delta_s|$ and $|\mu|$, one can project the Hamiltonian to the lower subband and linearize the electron dispersion relation around Fermi points. The charges are then obtained according to Eq.~\eqref{eqn:charge2}. The results are confirmed in the numerical calculation given in Fig.~\ref{nanowire} (a-b).

\textit{Tunneling transport spectroscopy -}
Now we propose to detect the ABSs with the tunneling transport spectroscopy. 
\begin{figure}[b]
\includegraphics[width=0.8\columnwidth]{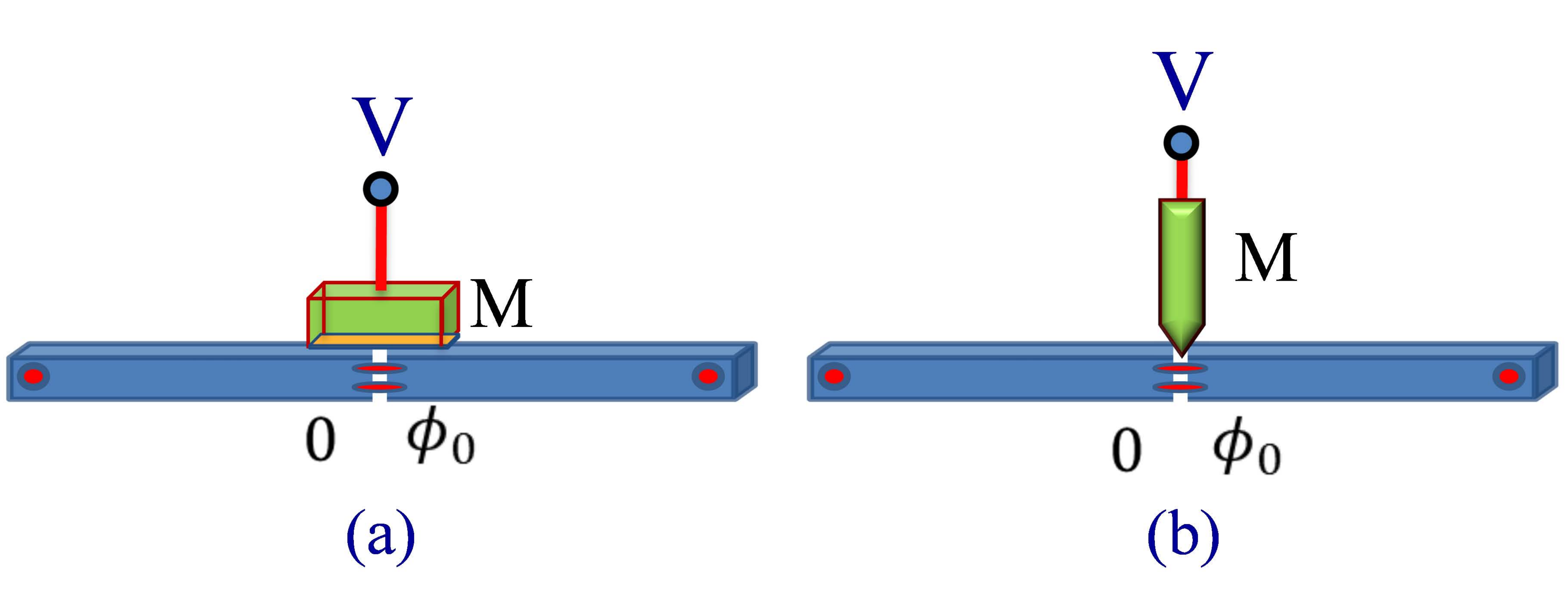}
\caption{(Color online) Sketch of the tunneling charge transport between the NM lead and the Josephson junction. A wide contact (a) and a narrow point contact (b) between the NM lead and the SC are considered.} \label{scheme}
\end{figure}
A single normal metallic (NM) lead with voltage $eV$ is coupled to the ABSs at the Josephson junction (Fig.~\ref{scheme}) in the tunneling regime. The coupling Hamiltonian is described by $H_T=\int dx t^*(x)d^\dag(x) \hat c(x)+h.c.$, where $d(x)$ is the electron annihilation operator in NM lead, and the tunneling coefficient $|t(x)|>0$ in the region $|x|<d_n/2$ with $d_n$ the width of the NM lead/SC contact region. Projecting the operator $c(x)$ onto the manifold of ABSs yields the effective tunneling Hamiltonian
\begin{eqnarray}\label{eqn:tunnel2}
H_T=\sum_{k,\nu}\sum_{\sigma=\pm}(f^{*}_{k,\nu\sigma}d_{k}^\dag-g_{k,\nu\sigma} d_{k}) b_{\nu\sigma}+h.c.,
\end{eqnarray}
where $f^{*}_{k,\nu\pm}=\int dx t_k^*(x)u_{\nu\pm}^{*}(x)e^{-ik_{R/L}x}$ and $g_{k,\nu\pm}=\int dx t_{k}(x)v_{\nu\pm}^{*}(x)e^{-ik_{L/R}x}$, with $t_{k}(x)=t(x)e^{ikx}$.
The tunneling current is calculated by the rate of change of the electron number $N=\sum_{k}d_{k}^\dag d_{k}$ in the NM lead: $I=-e\dot{N}=-\frac{ie}{\hbar}[H_T,N]$, which can be studied with the Keldysh formalism. Then we define the Keldysh contour Green's functions for ABSs by $Q_{\mu\mu'}(\tau,\tau')=-i\langle T_{K}[ b_{\mu}(\tau) b_{\mu'}^{\dag}(\tau')]\rangle$, with the index $\mu=(\nu,\pm)$, and the free lead electron Green's functions by $G_{k}^0(\tau,\tau')=-i\langle T_{K}[d_{k}(\tau) d_{k}^{\dag}(\tau')]\rangle_0$ and
${\bar G}^0_{k}(\tau,\tau')=-i\langle T_{K}[d_{k}^\dag(\tau) d_{k}(\tau')]\rangle_0$. Here $\langle\cdot\rangle_0$ represents the situation with $H_T=0$. The current is obtained by
\begin{eqnarray}\label{eqn:current1}
I&=&\frac{e}{\hbar}\int\frac{d\omega}{2\pi}\mbox{Tr}\bigr[\bigr(\bold\Sigma_{1}^{(e)}\bold Q-\bold Q\bold\Sigma_{1}^{(e)}\bigr)^<_\omega\bigr]+\nonumber\\
&&+\frac{e}{\hbar}\int\frac{d\omega}{2\pi}\mbox{Tr}\bigr[\bigr(\bold\Sigma_{2}^{(h)}\bold Q-\bold Q\bold\Sigma_{2}^{(h)}\bigr)^<_\omega\bigr],
\end{eqnarray}
where the trace is performed in the space spanned by $b_{\mu}$ modes. The self-energy matrices in real time space $\Sigma_{\mu\mu',1}^{(e)}(\tau,\tau')=\sum_{k}f_{k,\mu}f^{*}_{k,\mu'}G_{k}^0(\tau,\tau')$ and $
\Sigma_{\mu\mu',2}^{(h)}(\tau,\tau')=\sum_{k}g^{*}_{k,\mu}g_{k,\mu'}\bar{G}_{k}^0(\tau,\tau')$.
We derive the equation of motion for $Q_{\mu\mu'}(\tau,\tau')$ and obtain
$(i\partial_\tau-{\cal\bold E}^{\rm diag}-\bold \Sigma)\bold Q=1$,
where $\bold \Sigma=\bold\Sigma_{1}^{(e)}+\bold\Sigma_{2}^{(h)}$ and ${\cal\bold E}^{\rm diag}=\mbox{diag}\{...,{\cal E}_\mu,...\}$. The solution reads
$\bold Q=\bold Q^{0}+\bold Q^{0}\bold \Sigma\bold Q$, with $\bold Q^{0}=(\omega-{\cal\bold E}^{\rm diag})^{-1}$. 

For the NM lead, we consider the wide band limit that the transition matrix elements $f_{k,\mu'}$ and $g_{k,\mu'}$ are weakly energy dependent \cite{Meir}. In this case the self-energies are purely imaginary and we denote the retarded components by $\Sigma_{\mu\mu',1}^{(e)R}(\omega)=\frac{i}{2}\Upsilon_{\mu\mu',1}(\omega)$ and $\Sigma_{\mu\mu',2}^{(h)R}(\omega)=\frac{i}{2}\Upsilon_{\mu\mu',2}(-\omega)$.
Physically $\Upsilon_{\mu\mu',1}$ and $\Upsilon_{\mu\mu',2}$ determine the tunneling transmission probabilities for electrons and holes, respectively.

It is interesting to consider two different geometries of the NM lead characterized by the width $d_{n}$. First, we study the wide contact regime (Fig.~\ref{scheme} (a)) with the width $d_{n}$ greater than the localization length $l_{\rm ABS}$ of ABSs. Note that the Fermi wavelength in NM lead is much less than $l_{\rm ABS}$. This leads to a fast oscillating behavior for the phases of $f_{k,\mu}$ ($g_{k,\mu}$) versus $k$ \cite{Flensberg}. The off-diagonal elements of the self energy then vanishes, and $\Upsilon_{\mu\mu',1}\approx\delta_{\mu\mu'}\sum_{k}\int dxdx't_k^*(x)t_k(x')u^*_{\mu}(x)u_{\mu}(x')\delta(\epsilon_k-\omega)$ and $\Upsilon_{\mu\mu',2}\approx\delta_{\mu\mu'}\sum_{k}\int dxdx't_k^*(x)t_k(x')v^*_{\mu}(x')v_{\mu}(x)\delta(\epsilon_k-\omega)$, with $\epsilon_{k}$ the free electron energy in the NM lead.
The retarded Green's function for ABSs is given by $(Q^{R})_{\mu\mu}^{-1}(\omega)=\omega-{\cal E}_{\mu}+i\Upsilon_\mu$,
with $\Upsilon_\mu=(\Upsilon_{\mu\mu,1}+\Upsilon_{\mu\mu,2})/2$.
Bearing these results in mind, we obtain
the DTC at zero temperature
\begin{eqnarray}\label{eqn:conductance2}
\frac{dI}{dV}=\frac{2e^2}{\hbar}\sum_{\mu}\frac{\Upsilon_{\mu\mu,1}\Upsilon_{\mu\mu,2}}{(eV-{\cal E}_{\mu})^2+\Upsilon_\mu^2}.
\end{eqnarray}
One can verify with Eq.~\eqref{eqn:staterelation3} that $\Upsilon_{\mu\mu,1}/\Upsilon_{\mu\mu,2}=\gamma_\mu^2$.
Let ${\cal E}_{\rm min}$ be the minimum energy spacing for the ABSs, which depends on $\phi_0$ (Fig.~\ref{charge}(d)). For $\Upsilon_\mu^2\ll{\cal E}_{\rm min}^2$, the DTC has peaks at $eV_m={\cal E}_{\mu}$, corresponding to the resonant Andreev reflection induced by the ABSs. The resonant DTC for the peak values takes a simple form
\begin{eqnarray}\label{eqn:conductance3}
\biggr(\frac{dI}{dV}\biggr)_m=\frac{2e^2}{h}-\frac{2e^{*2}_\pm}{h},
\end{eqnarray}
which measures the charge $e^*_\pm$ carried by ABSs.

Second, for the narrow point contact regime with $d_{n}\ll l_{\rm ABS}$, the coupling can be approximated as $t(x)\approx t_0d_{n}\delta(x)$, which follows then $\Upsilon_{\mu\mu',1}\approx u^*_{\mu}(0)u_{\mu'}(0)|t_0|^2d_{n}^2N(\omega)$ and $\Upsilon_{\mu\mu',2}\approx v^*_{\mu}(0)v_{\mu'}(0)|t_0|^2d_{n}^2N(\omega)$, with $N(\omega)$ the density of states in NM lead. The main difference in the present regime is that the off-diagonal elements $\Upsilon_{\mu\mu'}$ ($\mu\neq\mu'$) are generically nonzero, which may lead to interference effect in the tunneling transports through different ABSs. Nevertheless, when $\Upsilon_{\mu\mu'}^2\ll{\cal E}_{\rm min}^2$, the off-diagonal terms are not important for the resonant Andreev reflection and we show that peak values of the DTC have the same form as in wide contact regime. Therefore Eq.~\eqref{eqn:conductance3} gives the generic formula of the DTC through resonant Andreev reflection induced by $each$ ABS with effective charge $e^*$.

\begin{figure}[ht]
\includegraphics[width=1.0\columnwidth]{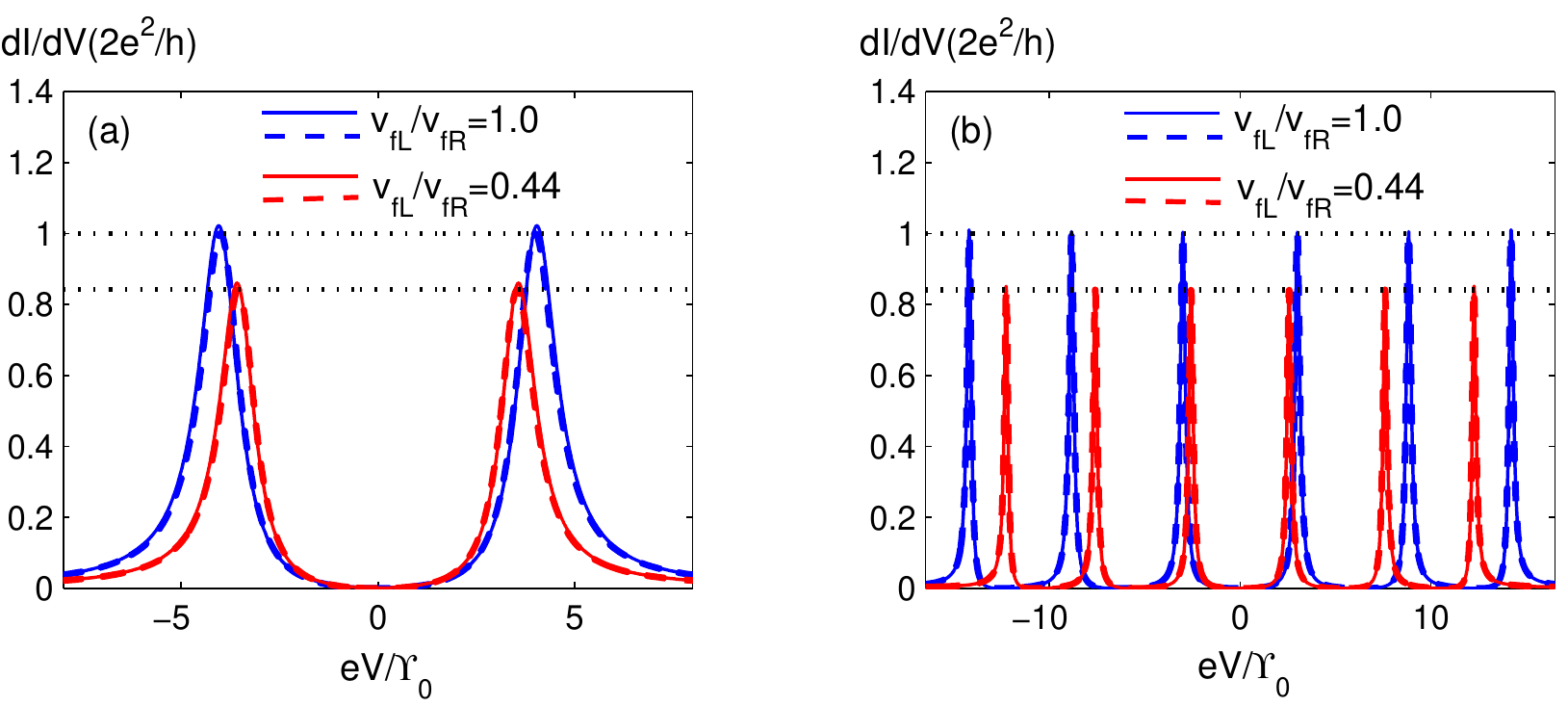}
\caption{(Color online) DTC $dI/dV$ (in units of $2e^2/h$) versus the bias $eV$ (in units of $\Upsilon_{0}$). The blue and red curves respectively correspond to the inversion symmetric case ($v_{fL}=v_{fR}$) and the case of $v_{fL}/v_{fR}=0.44$ ($t_2=0.2t_1$ and $\theta_0=\pi/2$ in lattice model). (a) The Josephson junction of a short length ($l<v_{fL}/|\Delta|$) is considered, with only two ABSs localized on the junction; (b) There are six ABSs localized on a long junction ($l\sim3v_{fR}/|\Delta|$). The equal height of peaks tells that the absolute value of the charge are the same for different ABSs in each parameter regime. The black dotted lines correspond to $(dI/dV)_m=2(e^2-e_\pm^{*2})/h$.}\label{fig_2}
\end{figure}
Fig.~\ref{fig_2} (a-b) shows the DTC as a function of bias $eV$. The peak values of the DTC at $eV_m={\cal E}_\mu$
confirm the results in Eq.~\ref{eqn:conductance3} for both the wide contact regime (solid curves) and narrow point contact regime (dashed curves).
In particular, for $v_{fR}=v_{fL}$ (blue curves) one has $e^*_\pm=0$ and the peak value of the DTC at resonant Andreev reflection is given by $2e^2/h$.
Physically, this can be understood that an incident electron leads to the unity reflection of a hole at the resonant Andreev reflection induced by a neutral ABS. This process is the same as the MF induced resonant Andreev
reflection (MIRAR) \cite{Law}. Our results show that the induced peak value $2e^2/h$ of the DTC
is {\em universal} for any {\em single neutral} ABS in 1D SCs.

In summary, we have presented a profound relation between the charge of ABSs and SIS in 1D topological SCs with small SC gap. A hidden symmetry is revealed for BdG equations, which protects that the charge of ABSs is solely determined by the Fermi velocities, regardless of other details. The charge of the ABSs can be measured by tunneling transport spectroscopy. Particularly, for system with SIS, the ABSs are charge neutral and each induces a DTC peak of height $2e^2/h$ at the resonant Andreev reflection with finite bias, similar to the cases for multiple MFs \cite{Law,Flensberg}.

We thank K. Sun, R. M. Lutchyn, T. D. Stanescu, K. T. Law, V. Yakovenko, Z. X. Liu, and X. Liu for helpful discussions. This work is supported by funding from  JQI-NSF-PFC, Microsoft-Q, DARPA, and QUEST.


\noindent

\newpage
\begin{widetext}

\begin{appendix}

\section{Andreev Bound states in One Dimensional Superconductor -- Supplementary Material}

In this supplementary material we provide the details of some results in the main text.

\section{Proof of the charge carried by Andreev bound states}

The BdG equations of $u_{\nu+}(x)$ and $v_{\nu+}(x)$ are given by $\frac{d}{dx}u_{\nu+}(x)=-i\frac{\tilde{\Delta}(x)}{v_{fR}}v_{\nu+}(x)+i\frac{{\cal E}_{\nu+}}{v_{fR}} u_{\nu+}(x)$ and $\frac{d}{dx}v_{\nu+}(x)=i\frac{\tilde{\Delta}^*(x)}{v_{fL}}u_{\nu+}(x)-i\frac{{\cal E}_{\nu+}}{v_{fL}}v_{\nu+}(x)$, from which we get further their complex conjugate counterparts by $\frac{d}{dx}u^*_{\nu+}(x)=i\frac{\tilde{\Delta}^*(x)}{v_{fR}}v^*_{\nu+}(x)-i\frac{{\cal E}_{\nu+}}{v_{fR}} u^*_{\nu+}(x)$ and $\frac{d}{dx}v^*_{\nu+}(x)=-i\frac{\tilde{\Delta}(x)}{v_{fL}}u^*_{\nu+}(x)+i\frac{{\cal E}_{\nu+}}{v_{fL}}v^*_{\nu+}(x)$.
With these formulas we can derive that
\begin{eqnarray}\label{eqn:ageneral6}
\frac{dG_{\nu+}(x)}{dx}=2\bigr(\frac{|v_{\nu+}(x)|^2}{v_{fR}}-\frac{|u_{\nu+}(x)|^2}{v_{fL}}\bigr)\frac{\mbox{Im}(\tilde{\Delta} u^*_{\nu+} v_{\nu+})}{|v|^2}.
\end{eqnarray}
It can also be verified from the BdG equations that $\frac{d}{dx}(v_{fR}|u_{\nu+}|^2-v_{fL}|v_{\nu+}|^2)=0$, which suggests that
\begin{eqnarray}\label{eqn:aproperty1}
f_{\nu+}(\Delta_1,\Delta_2)=v_{fR}|u_{\nu+}|^2-v_{fL}|v_{\nu+}|^2
\end{eqnarray}
is a function independent of the position.
The value of $f(\Delta_1,\Delta_2)$ can be determined by examining the asymptotic behavior of the ABS wave function. Note that the asymptotic behavior of $u_{\nu+}(x)$ and $v_{\nu+}(x)$ can be described by $u_{\nu+}(x\rightarrow+\infty)\sim\frac{A_1}{x^{\delta_1}}e^{-\gamma_1x+i\tilde{\gamma}_1x}$ and $v_{\nu+}(x\rightarrow+\infty)\sim\frac{A_2}{x^{\delta_2}}e^{-\gamma_2x+i\tilde{\gamma}_2x}$, respectively. Here $A_{1,2}, \delta_{1,2}, \gamma_{1,2}$, and $\tilde{\gamma}_{1,2}$ are constants. For $x\rightarrow+\infty$ we take that $\tilde{\Delta}(x\rightarrow+\infty)=\tilde{\Delta}_0$ is finite, and from BdG equations we have
\begin{eqnarray}\label{eqn:aproperty9}
(-\gamma_j+i\tilde{\gamma}_j)\frac{A_j}{x^{\delta_j}}e^{-\gamma_jx+i\tilde{\gamma}_jx}\mp i\frac{{\cal E}_{\nu+}}{\varpi_j}\frac{A_j}{x^{\delta_j}}e^{-\gamma_jx+i\tilde{\gamma}_jx}=-i\frac{\tilde{\Delta}_{0}}{\varpi_j}\frac{A_{j'}}{x^{\delta_{j'}}}e^{-\gamma_{j'}x+i\tilde{\gamma}_{j'}x},
\end{eqnarray}
where the higher-order infinitesimal terms are neglected,  $j\neq j'=1,2$, $\varpi_1=v_{fR}$, and $\varpi_2=v_{fL}$.
Comparing both sides of the above equations we observe that the exponents must satisfy the relations $\delta_1=\delta_2$ and $\gamma_1=\gamma_2$.
This confirms that $G_{\nu+}(x\rightarrow+\infty)$ is a constant, and thus $\frac{dG_{\nu+}(x\rightarrow+\infty)}{dx}\equiv0$. According to Eq. (\ref{eqn:ageneral6}) it follows that $G_{\nu+}(x\rightarrow\infty)=|u_{\nu+}(x)/v_{\nu+}(x)|^2=v_{fL}/v_{fR}$. Together with Eq. (\ref{eqn:aproperty1}) we have  $f_{\nu+}(\Delta_1,\Delta_2)=0$, which is valid for the entire position space and finally we reach $v_{fR}|u_{\nu+}(x)|^2=v_{fL}|v_{\nu+}(x)|^2$, completing the proof.

\section{The hidden symmetry of the BdG Hamiltonians ${\cal H}_\pm$}

\subsection{A. Non-degeneracy of the bound states for ${\cal H}_+$ (${\cal H}_-$)}

Before turning to the hidden symmetry, we show a basic property that the bound states of ${\cal H}_+$ (the same for ${\cal H}_-$) are non-degenerate. Let $\Phi_{1}=[u_{1}(x), v_{1}(x)]^T$ and $\Phi_{2}=[u_{2}(x), v_{2}(x)]^T$ be two degenerate bound states of ${\cal H}_+$ with energy ${\cal E}$. Then we have $\frac{d}{dx}u_{j}(x)=-i\frac{\tilde{\Delta}(x)}{v_{fR}}v_{j}(x)+i\frac{{\cal E}}{v_{fR}} u_{j}(x)$ and $\frac{d}{dx}v_{j}(x)=i\frac{\tilde{\Delta}(x)}{v_{fL}}u_{j}(x)-i\frac{{\cal E}}{v_{fL}}v_{j}(x)$ with $j=1,2$. From the BdG equations one can show that
\begin{eqnarray}\label{eqn:degeracy1}
\frac{d}{d x}(u_1v_2-u_2v_1)=i{\cal E}(\frac{1}{v_{fR}}-\frac{1}{v_{fL}})(u_1v_2-u_2v_1),
\end{eqnarray}
which has the solution $u_1v_2-u_2v_1=Ce^{i{\cal E}(\frac{1}{v_{fR}}-\frac{1}{v_{fL}})x}$ with $C$ a constant. Since $\Phi_{1,2}(x)$ are bound states, we have $u_1v_2-u_2v_1\rightarrow0$ for $x\rightarrow\infty$, which implies that $C=0$. Then one gets
\begin{eqnarray}\label{eqn:degeracy2}
\frac{u_1(x)}{v_1(x)}=\frac{u_2(x)}{v_2(x)}.
\end{eqnarray}
Namely, $\Phi_{1}(x)$ and $\Phi_{2}(x)$ are the same state, and therefore the bound state spectrum of ${\cal H}_+$ is non-degenerate.

\subsection{B. Hidden symmetry}

We consider ${\cal H}_+$ first. Let $\Phi_{\nu+}=[u_{\nu+}(x), v_{\nu+}(x)]^T$ satisfy ${\cal H}_+\Phi_{\nu+}={\cal E}_{\nu+}\Phi_{\nu+}$.
First, we apply the transformation
\begin{eqnarray}\label{eqn:unitary1}
\Phi_{\nu+}=(\mathcal{U}_{{\cal E}_{\nu+}}\mathcal{K})\tilde{\Phi}^*_{\nu+},
\end{eqnarray}
where $\mathcal{U}_{{\cal E}_{\nu+}}=e^{i{\cal E}_{\nu+}(\frac{1}{v_{fR}}-\frac{1}{v_{fL}})x}$ is the U(1) transformation and $\mathcal{K}$ is the complex conjugate operator. This leads to
\begin{eqnarray}\label{eqn:general4}
\frac{d}{dx}\tilde{u}^*_{\nu+}(x)&=&i\frac{\tilde{\Delta}^*(x)}{v_{fR}}\tilde{v}^*_{\nu+}(x)-i\frac{{\cal E}_{\nu+}}{v_{fL}} \tilde{u}^*_{\nu+}(x),
\end{eqnarray}
\begin{eqnarray}\label{eqn:general4'}
\frac{d}{dx}\tilde{v}^*_{\nu+}(x)&=&-i\frac{\tilde{\Delta}(x)}{v_{fL}}\tilde{u}^*_{\nu+}(x)+i\frac{{\cal E}_{\nu+}}{v_{fR}}\tilde{v}^*_{\nu+}(x).
\end{eqnarray}
Furthermore, we define the operator ${\cal P}_+=\tau_x\mathcal{L}_+$, where $\mathcal{L}_+=e^{\lambda_+\tau_z}$ is a Lorentz boost to rescale $u_{\nu+}(x)$ and $v_{\nu+}(x)$, with $\lambda_+=\ln(v_{fL}/v_{fR})^{1/2}$, and $\tau_{x,y,z}$ are the Pauli matrices acting on Nambu space. The operator ${\cal P}_+$ takes the form
\begin{eqnarray}\label{eqn:unitary3}
{\cal P}_+={\left[
\begin{matrix}
0 & e^{\lambda_+}\\
e^{-\lambda_+} & 0\end{matrix} \right]}=e^{\lambda_+}\tau_++e^{-\lambda_+}\tau_-,
\end{eqnarray}
Under the transformation
\begin{eqnarray}\label{eqn:unitary4}
\tilde{\Phi}^*_{\nu+}={\cal P}_+\tilde{\tilde{\Phi}}^*_{\nu+},
\end{eqnarray}
which gives $\tilde{\tilde{\Phi}}^*_{\nu+}=[\tilde{\tilde{v}}^*_{\nu+}(x),\tilde{\tilde{u}}^*_{\nu+}(x)]^T$, we obtain
\begin{eqnarray}\label{eqn:general4}
\frac{d}{dx}\tilde{v}^*_{\nu+}(x)&=&-i\frac{\tilde{\Delta}^*(x)}{v_{fR}}\tilde{\tilde{u}}^*_{\nu+}(x)+i\frac{{\cal E}_{\nu+}}{v_{fR}}\tilde{\tilde{v}}^*_{\nu+}(x),
\end{eqnarray}
\begin{eqnarray}\label{eqn:general4'}
\frac{d}{dx}\tilde{u}^*_{\nu+}(x)&=&i\frac{\tilde{\Delta}(x)}{v_{fL}}\tilde{\tilde{v}}^*_{\nu+}(x)-i\frac{{\cal E}_{\nu+}}{v_{fL}}\tilde{\tilde{u}}^*_{\nu+}(x).
\end{eqnarray}
The above equations can be rewritten as ${\cal H}_+\tilde{\tilde{\Phi}}^*_{\nu+}={\cal E}_{\nu+}\tilde{\tilde{\Phi}}^*_{\nu+}$. Therefore $\tilde{\tilde{\Phi}}^*_{\nu+}$ is still the eigenstate of ${\cal H}_+$ with the energy ${\cal E}_{\nu+}$. Note the {\em bound states} of ${\cal H}_+$ are {\em non-degenerate}. This leads to $\Phi_{\nu+}=e^{i\eta_+}\tilde{\tilde{\Phi}}_{\nu+}^*$, with $\eta_+$ arbitrary constant, which gives
\begin{eqnarray}\label{eqn:astaterelation3}
u_{\nu+}(x)=e^{\lambda_+} v_{\nu+}^{*}(x)e^{i\eta}e^{i{\cal E}_{\nu+}(\frac{1}{v_{fR}}-\frac{1}{v_{fL}})x}.
\end{eqnarray}

Similarly, for the Hamiltonian ${\cal H}_-(x)$, we introduce the similar transformations ${\cal P}_-, K$, and $U_{{\cal E}_{\nu-}}$, where $U_{{\cal E}_{\nu-}}=e^{i{\cal E}_{\nu-}(\frac{1}{v_{fR}}-\frac{1}{v_{fL}})x}$ and ${\cal P}_-=\tau_x\mathcal{L}_-$ with $\mathcal{L}_-=e^{\lambda_-\tau_z}$ and $\lambda_-=-\ln(v_{fL}/v_{fR})^{1/2}$. Under these transformations we also find ${\cal H}_-\tilde{\tilde{\Phi}}^*_{\nu-}={\cal E}_{\nu-}\tilde{\tilde{\Phi}}^*_{\nu-}$, which leads to $\Phi_{\nu-}=e^{i\eta_-}\tilde{\tilde{\Phi}}_{\nu-}^*$, and therefore
\begin{eqnarray}\label{eqn:astaterelation4}
u_{\nu-}(x)=e^{\lambda_-} v_{\nu-}^{*}(x)e^{i\eta_-}e^{i{\cal E}_{\nu-}(\frac{1}{v_{fR}}-\frac{1}{v_{fL}})x}.
\end{eqnarray}

The above consecutive transformations for ${\cal H}_\pm(x)$ can be summarized that under the transformation
\begin{eqnarray}\label{eqn:combineunitary1}
\Phi_{\nu\pm}(x)=(\mathcal{U}_{{\cal E}_{\nu\pm}}\mathcal{K}\tau_x{\cal L}_\pm)\tilde{\tilde{\Phi}}^*_{\nu\pm}(x),
\end{eqnarray}
the Hamiltonians are invariant
\begin{eqnarray}\label{eqn:combineunitary2}
(\mathcal{U}_{{\cal E}_{\nu\pm}}\mathcal{K}\tau_x{\cal L}_\pm)^{-1}{\cal H}_\pm(\mathcal{U}_{{\cal E}_{\nu\pm}}\mathcal{K}\tau_x{\cal L}_\pm)={\cal H}_\pm.
\end{eqnarray}
Therefore from Eqs. (\ref{eqn:astaterelation3}-\ref{eqn:astaterelation4}) we have
$\frac{|u_{\nu+}(x)|^2}{|v_{\nu+}(x)|^2}=e^{2\lambda_+}=\frac{v_{fL}}{v_{fR}}$ and $ \frac{|u_{\nu-}(x)|^2}{|v_{\nu-}(x)|^2}=e^{2\lambda_-}=\frac{v_{fR}}{v_{fL}}$.
This explains why $|u_{\nu\pm}(x)|^2$ and $|v_{\nu\pm}(x)|^2$ have such a simple relation and the charge only depends on the Fermi velocities for the linearized BdG Hamiltonians.

\section{Differential tunneling conductance}

From the formula $I=-\frac{ie}{\hbar}[H_T,N]$, we obtain the tunneling current by
\begin{eqnarray}\label{eqn:current1}
I=\frac{e}{\hbar}\sum_{k}\sum_{\mu=\pm}[f^{*}_{k,\mu}G_{\mu k}^{<}(0,0)-f_{k,\mu}G_{k\mu}^{<}(0,0)-g_{k,\mu} {\cal G}_{\mu k}^{<}(0,0)+g^{*}_{k,\mu}{\cal\bar G}_{k\mu}^{<}(0,0)],
\end{eqnarray}
where the mixed Green's functions are defined by $G_{k\mu}(\tau,\tau')=-i\langle T_{K}[d_{k}(\tau) b_{\mu}^{\dag}(\tau')]\rangle,
G_{\mu k}(\tau,\tau')=-i\langle T_{K}[b_{\mu}(\tau)d_{k}^\dag(\tau')]\rangle$,
${\cal G}_{\mu k}(\tau,\tau')=-i\langle T_{K}[b_{\mu}(\tau)d_{k}(\tau')]\rangle$, and
${\cal\bar G}_{k\mu}(\tau,\tau')=-i\langle T_{K}[d_{k}^\dag(\tau) b_{\mu}^{\dag}(\tau')]\rangle$. In the first order approximation we obtain for the lesser Green's functions that
\begin{eqnarray}\label{eqn:current2}
G_{k\mu}^{<}(\tau,\tau')&=&\sum_{\mu'}\int d\tau''\bigr[G_{k}^{0}(\tau,\tau'')f^{*}_{k,\mu'}Q_{\mu'\mu}(\tau'',\tau')\bigr]^{<},\\
G_{\mu k}^{<}(\tau,\tau')&=&\sum_{\mu'}\int d\tau''\bigr[f^{}_{k,\mu'}Q_{\mu\mu'}(\tau,\tau'')G_{k}^{0}(\tau'',\tau')\bigr]^{<},
\end{eqnarray}
and
\begin{eqnarray}\label{eqn:current2'}
{\cal G}_{\mu k}^{<}(\tau,\tau')&=&-\sum_{\mu'}\int d\tau''\bigr[g^{}_{k,\mu'}Q_{\mu\mu'}(\tau,\tau'')\bar{G}_{k}^{0}(\tau'',\tau')\bigr]^{<},\\
{\cal\bar{G}}_{k\mu}^{<}(\tau,\tau')&=&-\sum_{\mu'}\int d\tau''\bigr[\bar{G}_{k}^{0}(\tau,\tau'')g^{*}_{k,\mu'}Q_{\mu'\mu}(\tau'',\tau')\bigr]^{<}.
\end{eqnarray}
The tunneling current then recasts into
\begin{eqnarray}\label{eqn:current1}
I&=&\frac{e}{\hbar}\sum_{\mu\mu'}\int d\tau\bigr[\Sigma_{\mu\mu',1}^{(e)}(0,\tau)Q_{\mu'\mu}(\tau,0)-Q_{\mu\mu'}(0,\tau)
\Sigma_{\mu'\mu,1}^{(e)}(\tau,0)+\nonumber\\
&&+Q_{\mu\mu'}(0,\tau)\Sigma_{\mu'\mu,2}^{(h)}(\tau,0)-\Sigma_{\mu\mu',2}^{(h)}(0,\tau)Q_{\mu'\mu}(\tau,0)\bigr]^<,
\end{eqnarray}
where $\Sigma_{\mu\mu',1}^{(e)}(\tau,\tau')=\sum_{k}f_{k,\mu}f^{*}_{k,\mu'}G_{k}^0(\tau,\tau')$ and $
\Sigma_{\mu\mu',2}^{h)}(\tau,\tau')=\sum_{k}g^{*}_{k,\mu}g_{k,\mu'}\bar{G}_{k}^0(\tau,\tau')$ are the corresponding self-energies.
The Eq. (\ref{eqn:current1}) can also be written as
$I=I_1+I_2$, where
\begin{eqnarray}\label{eqn:current5}
I_{1}&=&\frac{e}{\hbar}\int\frac{d\omega}{2\pi}\mbox{Tr}\bigr[\bigr(\bold\Sigma_{1}^{(e)}\bold Q-\bold Q\bold\Sigma_{1}^{(e)}\bigr)^<_\omega\bigr]\nonumber\\
&=&\frac{e}{\hbar}\int\frac{d\omega}{2\pi}\mbox{Tr}\bigr[\bigr(\bold Q^{R}_\omega-\bold Q^{A}_\omega\bigr)\bold\Sigma_{1\omega}^{(e)<}+\bold Q^{<}_\omega\bigr(\bold\Sigma_{1\omega}^{(e)A}-\bold\Sigma_{1\omega}^{(e)R}\bigr)\bigr],\\
I_{2}&=&\frac{e}{\hbar}\int\frac{d\omega}{2\pi}\mbox{Tr}\bigr[\bigr(\bold\Sigma_{2}^{(h)}\bold Q-\bold Q\bold\Sigma_{2}^{(h)}\bigr)^<_\omega\bigr]\nonumber\\
&=&\frac{e}{\hbar}\int\frac{d\omega}{2\pi}\mbox{Tr}\bigr[\bigr(\bold Q^{R}_\omega-\bold Q^{A}_\omega\bigr)\bold\Sigma_{2\omega}^{(h)<}+\bold Q^{<}_\omega\bigr(\bold\Sigma_{2\omega}^{(h)A}-\bold\Sigma_{2\omega}^{(h)R}\bigr)\bigr].
\end{eqnarray}
The Dyson equation of $Q_{\mu\mu'}(\tau,\tau')$ can be derived through $i\partial_\tau Q_{\mu\mu'}(\tau,\tau')=\delta_{\mu\mu'}\delta(\tau-\tau')+i\langle T_K\bigr([H, b_{\mu}](\tau) b_{\mu'}^\dag(\tau')\bigr)\rangle$, which follows that
$(i\partial_\tau-{\cal\bold E}^{\rm diag}-\bold \Sigma)\bold Q=1$.
Here $\bold \Sigma=\bold\Sigma_{1}^{(e)}+\bold\Sigma_{2}^{(h)}$ and ${\cal\bold E}^{\rm diag}=\mbox{diag}\{...,{\cal E}_\mu,...\}$ is the diagonal matrix composed of eigenvalues of the ABSs. The solution reads
$\bold Q=\bold Q^{0}+\bold Q^{0}\bold \Sigma\bold Q$, with $\bold Q^{0}=(\omega-{\cal\bold E}^{\rm diag})^{-1}$.

For the NM lead, we consider the wide band limit that the transition matrix elements $f_{k,\mu'}$ and $g_{k,\mu'}$ are weakly energy dependent \cite{aMeir}. In this case the self-energies are purely imaginary and the retarded components read $\Sigma_{\mu\mu',1}^{(eh)R}(\omega)=\frac{i}{2}\Upsilon_{\mu\mu',1}(\omega)$ and $\Sigma_{\mu\mu',2}^{(h)R}(\omega)=\frac{i}{2}\Upsilon_{\mu\mu',2}(-\omega)$,
where
$\Upsilon_{\mu\mu',1}(\omega)=2\pi\sum_{k}f_{k,\mu}f^{*}_{k,\mu'}\delta(\omega-\epsilon_{k})$ and
$\Upsilon_{\mu\mu',2}(\omega)=2\pi\sum_{k}g_{k,\mu}g^{*}_{k,\mu'}\delta(\omega-\epsilon_{k})$, with  $\epsilon_{k}$ the free electron energy in the NM lead. The lesser components $\Sigma_{\mu\mu',1}^{(e)<}(\omega)=i\Upsilon_{\mu\mu',1}(\omega)f(\omega-eV),
\Sigma_{\mu\mu',2}^{(h)<}(\omega)=i\Upsilon_{\mu\mu',2}(-\omega)[1-f(\omega-eV)]$, and $\bold Q^{<}_\omega=\bold Q^{R}_\omega\bold\Sigma_{\omega}^{<}\bold Q^{A}_\omega$. This leads to
\begin{eqnarray}\label{eqn:current7}
I_1&=&\frac{e}{\hbar}\int\frac{d\omega}{2\pi}\mbox{Tr}\bigr\{\bold Q^{R}(\omega)\bold\Upsilon_2(\omega)\bold Q^{A}(\omega)\bold\Upsilon_1(\omega)\bigr\}[1-f(\omega-eV)],\\
I_2&=&\frac{e}{\hbar}\int\frac{d\omega}{2\pi}\mbox{Tr}\bigr\{\bold Q^{R}(\omega)\bold\Upsilon_1(\omega)\bold Q^{A}(\omega)\bold\Upsilon_2(\omega)\bigr\}[1-f(\omega-eV)].
\end{eqnarray}
It can be verify that $I_1=I_2$. The differential conductance is then given by
\begin{eqnarray}\label{eqn:conductance1}
\frac{dI}{dV}=\frac{2e^2}{\hbar}\int\frac{d\omega}{2\pi}\mbox{Tr}\bigr\{\bold Q^{R}(eV)\bold \Upsilon_2\bold Q^{A}(eV)\bold \Upsilon_1\}\frac{d f(\omega-eV)}{d\omega}.
\end{eqnarray}

The Fermi wavelength
in the NM lead is much less than the ABS localization length $l_{\rm ABS}$, and also typically much less than $k_{R,L}^{-1}$ in nanowire systems. For the wide contact regime with the width $d_n$ of NM lead greater than $l_{\rm ABS}$, the functions $f_{k,\mu}$ ($g_{k,\mu}$) exhibit a fast phase variation versus $k$, and the off-diagonal elements of the self energy vanishes. We then reach that $\Upsilon_{\mu\mu',1}\approx\delta_{\mu\mu'}\sum_{k}\int dxdx't_k^*(x)t_k(x')u^*_{\mu}(x)u_{\mu}(x')\delta(\epsilon_k-\omega)$ and $\Upsilon_{\mu\mu',2}\approx\delta_{\mu\mu'}\sum_{k}\int dxdx't_k^*(x)t_k(x')v^*_{\mu}(x')v_{\mu}(x)\delta(\epsilon_k-\omega)$.
The retarded Green's function for ABSs $(Q^{R})_{\mu\mu}^{-1}(\omega)=\omega-{\cal E}_{\mu}+i\Upsilon_\mu$,
with $\Upsilon_\mu=(\Upsilon_{\mu\mu,1}+\Upsilon_{\mu\mu,2})/2$. With these results we obtain
the DTC at zero temperature
\begin{eqnarray}\label{eqn:conductance2}
\frac{dI}{dV}&=&\frac{2e^2}{h}\mbox{Tr}\bigr\{\bold Q^{R}(eV)\bold\Upsilon_2\bold Q^{A}(eV)\bold\Upsilon_1\}\nonumber\\
&=&\frac{2e^2}{\hbar}\sum_{\mu}\frac{\Upsilon_{\mu\mu,1}\Upsilon_{\mu\mu,2}}{(eV-{\cal E}_{\mu})^2+\Upsilon_\mu^2},
\end{eqnarray}
which is the Eq. (9) in the main text. For $\Upsilon_\mu^2\ll{\cal E}_{\rm min}^2$ with ${\cal E}_{\rm min}$ the minimum energy spacing for the ABSs, the DTC has peaks at $eV_m\approx\pm{\cal E}_\mu$, with the peak values given by $\bigr(\frac{dI}{dV}\bigr)_m=\frac{2e^2}{h}-\frac{2e^{*2}_\pm}{h}$,
which measures the charges $e^*_\pm$ carried by ABSs.


\noindent

\end{appendix}

\end{widetext}

\end{document}